\documentclass[pre,twocolumn]{revtex4}

\usepackage{amssymb}
\usepackage{amsmath}
\usepackage{graphicx}
\usepackage{epstopdf}
\usepackage{color}
\usepackage{hyperref}

\newcommand{\Ai}{\mathrm{Ai}}
\renewcommand{\P}{\mathrm{Prob}}
\newcommand{\rmd}{\mathrm{d}}

\begin{document}

\title[KPZ numerical]{The height distribution of the KPZ equation with sharp wedge initial condition: numerical evaluations}
\author{Sylvain Prolhac and Herbert Spohn}
\affiliation{Zentrum Mathematik and Physik Department,\\
Technische Universit\"at M\"unchen,\\
D-85747 Garching, Germany}
\date{\today}

\begin{abstract}
The time-dependent probability distribution function of the height for the Kardar-Parisi-Zhang equation with sharp wedge initial conditions has been obtained recently as a convolution between the Gumbel distribution and a difference of two Fredholm determinants. We evaluate numerically this distribution over the whole time span. The crossover from the short time behavior, which is Gaussian, to the long time behavior, which is governed by the GUE Tracy-Widom distribution, is clearly visible.
\end{abstract}

\pacs{05.20.-y 05.40.-a 05.70.Np}
\keywords{Kardar-Parisi-Zhang equation, height fluctuations, numerical evaluations of Fredholm determinants}

\maketitle


\section{Introduction}
\label{Section introduction}
A recent experiment \cite{TS10.1} on a film of turbulent liquid crystal realized the growth of a droplet made up of a stable phase and expanding into a metastable phase. The shape fluctuations of the droplet were measured with high precision. In approximation the top part of this droplet can be represented by a height function $h(x,t)$, $x$ on the real line, and its dynamics is governed the one-dimensional version of the Kardar-Parisi-Zhang (KPZ) equation
\begin{equation}
\label{KPZ}
\partial_{t}h(x,t)=\tfrac{1}{2}\lambda(\partial_{x}h(x,t))^{2}+\nu\,\partial_{x}^{2}h(x,t)+\sqrt{D}\,\eta(x,t)\,
\end{equation}
\cite{KPZ86.1}. The first term on the right hand side is a non-linearity which results from growth normal to the interface at leading order in the local slope. The second term is the surface tension which tends to smoothen the interface. The last term describes the random nucleation events at the interface and is modeled by a Gaussian white noise $\eta$ with covariance
\begin{equation}
\label{<etaeta>}
\langle\eta(x,t)\eta(x',t')\rangle=\delta(x-x')\delta(t-t')\,.
\end{equation}
By scale invariance of the KPZ equation one can switch to dimensionless variables. We make the choice
\begin{equation}
\lambda = 1\,,\quad \nu = 1/2\,,\quad D = 1\,.
\end{equation}
The conversion to a general set of coefficients is discussed in \cite{PS11.1}, for example. We will consider the case of sharp wedge initial conditions,
\begin{equation}
\label{KPZ t=0}
h(x,0)=-|x|/\varepsilon\, \quad \mathrm{with} \quad \varepsilon > 0\,,\, \varepsilon\to0\,.
\end{equation}
The solution to (\ref{KPZ}) -- (\ref{KPZ t=0}) is written as
\begin{equation}
\label{h[xi]}
h(x,t) = -\frac{t}{24} - \frac{x^2}{2t} + \Big(\frac{t}{2}\Big)^{1/3}\xi(x,t)\,.
\end{equation}
The flattening parabola should be viewed as the top part of the droplet in the experiment of \cite{TS10.1} and $\xi(x,t)$ are the superimposed fluctuations. Of course, the KPZ equation holds in greater generality, in particular also for interface motion and growth models in higher dimensions. For surveys on the earlier developments we refer to \cite{BS95.1,HHZ95.1,K07.1}. Recently the KPZ equation has been used as a challenging test ground for nonequilibrium RG techniques \cite{F06.1,CCDW10.1}.

The subtraction $x^2/2t$ is uniquely fixed by the requirement that $\langle\xi(x,t)\rangle$ is independent of $x$ for given $t>0$. In fact, by the scale invariance of the KPZ equation, for fixed $t$ also higher order correlations depend only on the relative distance in $x$. $\langle\xi(x,t)\rangle$ depends on $t$. To determine its value is less obvious, since the construction of the solution to (\ref{KPZ}) -- (\ref{KPZ t=0}) already requires a diverging uniform shift in the $h$-direction. This is most easily explained for the initial condition $h_\varepsilon(x,0)$. The construction of the solution requires
\begin{equation}
\lim_{\varepsilon \to 0} \mathrm{e}^{h_\varepsilon(x,0)} = \delta(x)\,,
\end{equation}
which means that $h_\varepsilon(x,0) = - \varepsilon^{-1}|x| - \log (2\varepsilon)$ with $\log (2\varepsilon)$ diverging as $\varepsilon \to 0$. Our convention here is to fix the first exponential moment of the solution to (\ref{KPZ}) -- (\ref{KPZ t=0}) as
\begin{equation}
\langle \mathrm{e}^{h(x,t)}\rangle = \frac{1}{\sqrt{2\pi t}}\,\mathrm{e}^{-x^2/2t}\,.
\end{equation}
Therefore
\begin{equation}
\langle \exp((t/2)^{1/3}\xi(x,t)) \rangle = \frac{1}{\sqrt{2\pi t}}\,\mathrm{e}^{t/24}\,.
\end{equation}
The rationale behind introducing the scale $(t/2)^{1/3}$ for $\xi(x,t)$ and subtracting $t/24$ results from the KPZ scaling theory which asserts that $\xi(x,t) = v_\infty t + \mathcal{O}(t^{1/3})$ for large $t$ with $v_\infty$ the true asymptotic velocity. $v_\infty$ is non-universal, $v_\infty = - 1/24$ in our case and our units. The numerical factor $(1/2)^{1/3}$ will be discussed below.

As in the experiment \cite{TS10.1}, the quantity of prime physical interest is the fluctuating amplitude $\xi(x,t)$ at fixed $x,t$. By stationarity one may pick $x=0$ and we define
\begin{equation}
\xi_t = \xi(0,t)\,.
\end{equation}
To determine numerically the statistics of $\xi_t$ is extremely demanding and has never been accomplished, see \cite{KS04.1} for the best results available. Thus it was a great breakthrough to have an exact formula for the distribution function $ F_{t}(s) = \P\left(\xi_t < s\right) $ \cite{ACQ11.1,SS10.1,SS10.3,CLDR10.1,D10.1}. $F_{t}(s)$ can be written as the convolution of the Gumbel distribution with some function $g_{t}$,
\begin{equation}
\label{Ft(s)}
F_{t}(s)=1-\int_{-\infty}^{\infty}\rmd u\,\exp\left[(-\exp\big((t/2)^{1/3}(s-u)\big)\right]g_{t}(u)\,,
\end{equation}
where $g_{t}$ turns out to be given by the difference of two Fredholm determinants,
\begin{equation}
\label{gt}
g_{t}(u)=\det\big(\openone-P_{u}(B_{t}-P_{\Ai})P_{u}\big)-\det\big(\openone-P_{u}B_{t}P_{u}\big)\,.
\end{equation}
The operator $P_{u}$ is the projection on $[u,\infty)$, while $B_{t}$ and $P_{\Ai}$ are defined in terms of the Airy function $\Ai$ respectively by
\begin{align}
\label{Bt}
\langle z|B_{t}|z'\rangle=\int_{0}^{\infty}\rmd\lambda\,\bigg(&\frac{\Ai(z+\lambda)\Ai(z'+\lambda)}{1-\exp(-(t/2)^{1/3}\lambda)}\\
&+\frac{\Ai(z-\lambda)\Ai(z'-\lambda)}{1-\exp((t/2)^{1/3}\lambda)}\bigg)\,\nonumber
\end{align}
and
\begin{equation}
\langle z|P_{\Ai}|z'\rangle=\Ai(z)\Ai(z')\,.
\end{equation}
We recall that the Fredholm determinant of a well-behaved (``trace class'') integral operator with kernel $A$ can be defined by
\begin{align}
\label{def Fredholm det}
\det(\openone+A)&=1+\sum_{m=1}^{\infty}\frac{1}{m!}\\
&\times\int_{-\infty}^{\infty}\mathrm{d}y_{1}\,\ldots\,\mathrm{d}y_{m}\,\det\left(A(y_{j},y_{k})\right)_{j,k=1,\ldots,m}\,.\nonumber
\end{align}

While (\ref{Ft(s)}), (\ref{gt}) is decisive progress compared to a direct numerical simulation of the KPZ equation, it is still necessary to produce plots of the probability density function $F'_t(s)$, which is the main goal of our communication. The numerical task is demanding, since it concerns the computation of the Fredholm determinants, as will be explained in Section 2. For the data plot it is natural to distinguish short time, see Section 3, and long time behavior, see Section 4.


\section{Numerical evaluation of \texorpdfstring{$g_{t}$}{gt}}
To plot the distribution $F_{t}$, one needs to evaluate the $t$-dependent family of functions $g_{t}(u)$, see (\ref{gt}). Efficient, to some extent even optimal techniques to numerically compute Fredholm determinants have been developed recently \cite{B10.1,B10.2}. The basic idea is to employ a properly adapted discretization of the integrals appearing in the definition (\ref{def Fredholm det}) of the Fredholm determinant.

As an example, let us first consider the case of a kernel $A(y,y')$ with support in $[-1,1]^2$, so that all the integrals in (\ref{def Fredholm det}) range from $-1$ to $1$. $A$ is assumed to be smooth in both variables. Then an integral is discretized as
\begin{equation}
\label{discretization int}
\int_{-1}^{1}\mathrm{d}y\,f(y)\approx\sum_{\ell=1}^{n}w_{\ell}f(y_{\ell})\,,
\end{equation}
The particular choice of the weights $w_{\ell}$ and base points $y_{\ell}$, $\ell=1,\ldots,n$, is called a \textit{quadrature rule}. The simplest choice is the rectangular quadrature $w_{\ell}=2/n$, $y_{\ell}=-1+2\ell/n$. This choice, however, generically does not ensure the fast convergence of the discretized integrals. As emphasized in \cite{B10.1}, it is far better to use instead the Gauss-Legendre quadrature rule which leads to an exponentially fast convergence when $n\to\infty$. For the Gauss-Legendre quadrature, $y_{\ell}$ is the $\ell$-th zero of the $n$-th Legendre polynomial $P_{n}(y)=(2^{n}n!)^{-1}\partial_{y}^{n}(y^{2}-1)^{n}$ and $w_{\ell}=2/(nP_{n-1}(y_{\ell})P_{n}'(y_{\ell}))$. The Gauss-Legendre quadrature is singled out as the only quadrature rule which gives the exact result for the integral over the functions $f$ polynomial of degree at most $2n-1$. After discretization, the Fredholm determinant reduces to a single $n\times n$ determinant using the von Koch formula \cite{B10.1}
\begin{align}
&\det(\openone+A)\approx1+\sum_{m=1}^{\infty}\frac{1}{m!}\nonumber\\
&\times\sum_{\ell_{1},\ldots,\ell_{m}=1}^{n}w_{\ell_{1}}\ldots w_{\ell_{m}}\det\left(A(y_{\ell_{j}},y_{\ell_{k}})\right)_{j,k=1,\ldots,m}\nonumber\\
&=\det(\openone+\sqrt{w_{\ell}}A(y_{\ell},y_{\ell'})\sqrt{w_{\ell'}})_{\ell,\ell'=1,\ldots,n}\,.
\end{align}
\begin{figure}
\begin{center}\begin{picture}(0,0)\put(68,47){$g_{t}(u)$}\put(77,29){$u$}\end{picture}\includegraphics[width=80mm]{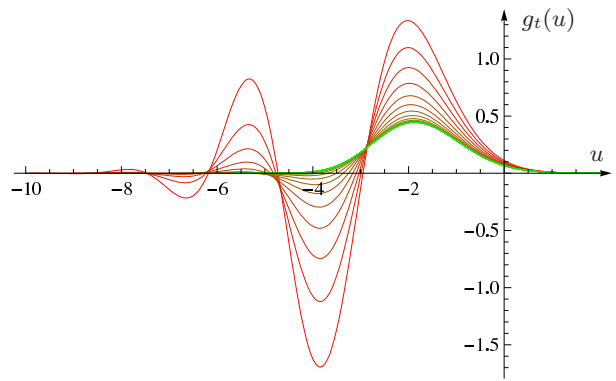}\end{center}
\caption{(Color online) $g_{t}(u)$ from short times (red, larger amplitudes) to long times (green, lower amplitudes) for the values of $t$ as listed in (\ref{list t}).}
\label{fig gt}
\end{figure}

Since the Fredholm determinants in (\ref{gt}) involve kernels with support on $\mathbb{R}^2$, and not on $[-1,1]^2$, a change of variables must be performed in the integrals, using a function $\varphi$ growing from $\varphi(-1)=-\infty$ to $\varphi(1)=\infty$,
\begin{equation}
\int_{-\infty}^{\infty}\mathrm{d}z\,f(z)=\int_{-1}^{1}\mathrm{d}y\,\varphi'(y)f(\varphi(y))\,.
\end{equation}
The Fredholm determinant is then approximated as
\begin{align}
\det(\openone+A)\approx&\det\big(\openone+\sqrt{w_{\ell}\varphi'(y_{\ell})}\\
&\times A(\varphi(y_{\ell}),\varphi(y_{\ell'}))\sqrt{w_{\ell'}\varphi'(y_{\ell'})}\big)_{\ell,\ell'=1,\ldots,n}\,,\nonumber
\end{align}
which converges to the true value of $\det(\openone+A)$ in the limit $n\to\infty$. If the kernel itself is given by an integral, like $B_{t}$ of (\ref{Bt}), this integral is also approximated by the Gauss-Legendre discretization (\ref{discretization int}) with the same value of $n$.

For the numerical computation of the Fredholm determinants in $g_{t}(u)$, we choose $n=30$ and $\varphi(y)=10\tan(\pi y/2)$. All calculations are done with double precision floating points. The values of $t$ are chosen so to achieve a good resolution. In all figures we plot the $19$ values
\begin{align}
\label{list t}
&t=0.25\hspace{5pt} 0.35\hspace{5pt} 0.5\hspace{5pt} 0.75\hspace{5pt} 1.2\hspace{5pt} 2 \hspace{5pt} 3.5\hspace{5pt} 6.5\hspace{5pt} 13\hspace{5pt} 25\hspace{5pt} 50 \nonumber\\
&\hspace{20pt}100 \hspace{5pt}250 \hspace{5pt}500\hspace{5pt} 1000\hspace{5pt} 2500\hspace{5pt} 5000\hspace{5pt} 10000\hspace{5pt} 20000\,.
\end{align}
For the final integration over $u$ in (\ref{Ft(s)}) we use an equally spaced grid with step size $\delta u=0.05$ starting from $u=0$ and stopping after $5$ consecutive values of $|g_{t}(u)|$ which are smaller than $10^{-5}$. For all values of $t$ considered, ranging from $0.25$ to $20000$, the upper bound $u_{\mathrm{max}}$ is around $u_{\mathrm{max}}=3.25$, while the lower bound $u_{\mathrm{min}}$ decreases towards short times ($u_{\mathrm{min}}=-5.65$, $-8$, $-10.9$ respectively for $t=20000$, $2$, $0.25$).

Since the convergence when $n\to\infty$ is slower for small values of $u$, the numerical computation of the distribution $F_{t}$ is more difficult at short times. With $n=30$, an accurate result cannot be achieved for times shorter than $t\approx0.25$.

The results for $g_{t}$ are displayed in Fig. \ref{fig gt}. Because of oscillations the function $g_{t}$ always takes some negative values: it is only the convolution with the Gumbel distribution which yields a \textit{bona fide} probability density function. The density $F_{t}'$ is plotted in Figs. \ref{fig Ft' shifted} and \ref{fig Ft'}.


\section{Short time behavior}
\begin{figure}
\begin{center}\begin{picture}(0,0)\put(36.5,49){$\tilde{F_{t}}'(s)$}\put(79.5,5){$s$}\end{picture}\includegraphics[width=80mm]{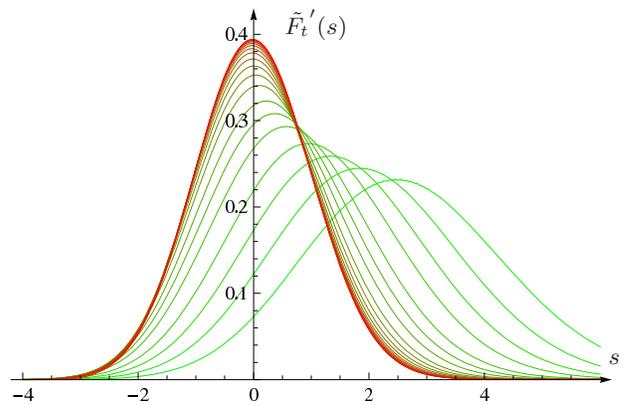}\end{center}
\caption{(Color online) Shifted and rescaled probability density $\tilde{F_{t}}'(s)=\sigma F_{t}'(\sigma s+\mu)$ from short times (red, leftmost curves) to long times (green, rightmost curves) for the values of $t$ as listed in (\ref{list t}).}
\label{fig Ft' shifted}
\end{figure}
\begin{figure}
\begin{center}\begin{picture}(0,0)\put(80,5){$t$}\end{picture}\includegraphics[width=80mm]{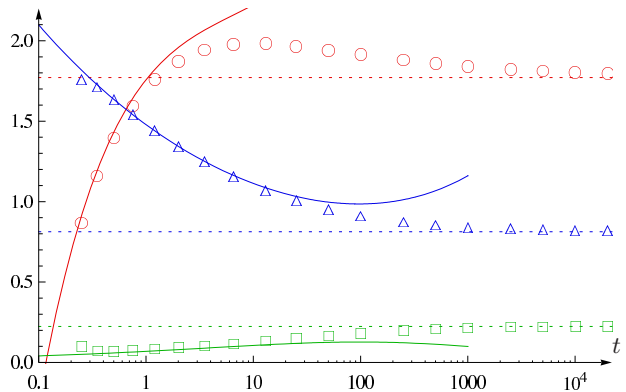}\end{center}
\caption{(Color online) First three normalized cumulants of the distribution $F_{t}$ as a function of $t$. Short time expansion of $-C_{1}$ (solid red line), $C_{2}$ (solid blue line), and $C_{3}/C_{2}^{3/2}$ (solid green line) as in (\ref{C}) and numerical evaluation of the cumulants $-C_{1}$ (red circles), $C_{2}$ (blue triangles), $C_{3}/C_{2}^{3/2}$ (green squares) for the values of $t$ as listed in (\ref{list t}). The dotted lines represent the asymptotic Tracy-Widom values (\ref{C TW 1}) -- (\ref{C TW 3}).}
\label{fig cumulants}
\end{figure}
For short times a convenient choice for the fluctuating amplitude is
\begin{equation}
h(0,t) = -\frac{t}{24} - \log \sqrt{2\pi t} + \Big(\frac{\pi t}{4}\Big)^{1/4}\zeta_t\,,
\end{equation}
compare with (\ref{h[xi]}). As established in \cite{ACQ11.1}, $\zeta_t$ converges for $t\to0$ to a standard Gaussian. A short time expansion for the cumulants of $\zeta_t$ is carried out in \cite{CLDR10.1}. The short $t$ numerical values are in perfect agreement with these results. In our display we focus on the intermediate time behavior for which, as before, we use the fluctuating amplitude $\xi_t$ with distribution $F_t(s)$. The properly scaled $F_{t}'(s)$ becomes Gaussian,
\begin{equation}
\label{Ft short time}
\lim_{t\to0}F_{t}(\sigma s+\mu)=\int_{-\infty}^{s}\mathrm{d}u\,\frac{\mathrm{e}^{-u^{2}/2}}{\sqrt{2\pi}}\,,
\end{equation}
where the parameters $\mu=-2^{-2/3}t^{-1/3}\log(2\pi t)-2^{-5/3}\pi^{1/2}t^{1/6}$ and $\sigma^{2}=2^{-1/3}\pi^{1/2}t^{-1/6}$ are the leading order of the mean and variance of the distribution $F_{t}$ when $t\to0$. This limit behavior is well reproduced in Fig. \ref{fig Ft' shifted}. 

From the evaluation of the distribution $F_{t}$, one can obtain the low order cumulants. The $j$-th moment $M_{j}$ is defined by
\begin{equation}
M_{j}=\int_{-\infty}^{\infty}\mathrm{d}s\,s^{j}F_{t}'(s-2^{-2/3}t^{-1/3}\log(2\pi t))\,.
\end{equation}
For all values of $t$ considered, $F_{t}(s)$ is concentrated in the interval $[-6,4]$. Therefore numerically the moments are approximated by the Riemann sum of $s^{j}F_{t}'(s)$ with step size $\delta s=0.05$ and $s$ ranging over this interval. The three first cumulants are given by
\begin{equation}
C_{1}=M_{1}\,,\quad C_{2}=M_{2}-M_{1}^{2}\,,\quad C_{3}=M_{3}-3M_{1}M_{2}+2M_{1}^{3}\,.
\end{equation}
We compare them to the short time expansion of \cite{CLDR10.1}, which with our conventions reads
\begin{align}
\label{C}
C_{1}&=-\sqrt{\frac{\pi}{8}}\,(t/2)^{1/6}-\left(\frac{1}{2}+\frac{3\pi}{8}-\frac{8\pi}{9\sqrt{3}}\right)(t/2)^{2/3}+\ldots\,,\nonumber\\
C_{2}&=\sqrt{\frac{\pi}{2}}\,(t/2)^{-1/6}+\left(1+\frac{5\pi}{4}-\frac{8\pi}{3\sqrt{3}}\right)(t/2)^{1/3}+\ldots\,,\nonumber\\
C_{3}&=\pi\left(\frac{8}{3\sqrt{3}}-\frac{3}{2}\right)+\ldots\,.
\end{align}
The comparison is presented in Fig. \ref{fig cumulants}. The fit is fairly accurate at short times. The discrepancy for the smallest values of $t$ signals that the discretization of the Fredholm determinant with $n=30$ is not sufficiently accurate for $t\approx0.25$.


\section{Long time behavior}
\begin{figure}
\begin{center}\begin{picture}(0,0)\put(42,48){$F_{t}'(s)$}\put(79,5){$s$}\end{picture}\includegraphics[width=80mm]{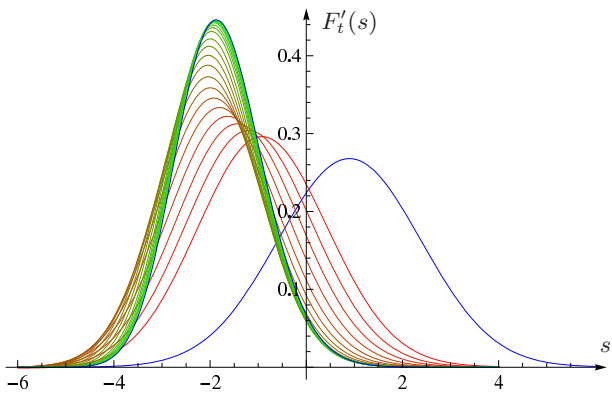}\end{center}
\caption{(Color online) Probability density function $F_{t}'(s)$ for time $t$ from short times (red, lower curves) to long times (green, upper curves) for the values of $t$ as listed in (\ref{list t}). For $t\to\infty$, the density converges to the GUE Tracy-Widom distribution $F_{\text{GUE}}$ (upper blue curve) and, for $t\to 0$, $F_{t}'(s)$ becomes a Gaussian (rightmost blue curve) with mean and variance increasing respectively as $t^{-1/3}\log t$ and $t^{-1/6}$.}
\label{fig Ft'}
\end{figure}
In the limit $t\to\infty$, $F_{t}(s)$ converges to the GUE Tracy-Widom distribution,
\begin{equation}
\label{Ft long time}
\lim_{t\to\infty}F_{t}(s)=F_{\text{GUE}}(s)=\det(1-P_{s}KP_{s})\,,
\end{equation}
where $K$ is the Airy kernel
\begin{align}
\label{K}
\langle z|K|z'\rangle&=\int_{0}^{\infty}\rmd\lambda\,\Ai(z+\lambda)\Ai(z'+\lambda)\nonumber\\
&=\frac{\Ai(z)\Ai'(z')-\Ai'(z)\Ai(z')}{z-z'}\,.
\end{align}
Just like in the case of a standard Gaussian, by convention the Tracy-Widom distribution is defined as in (\ref{Ft long time}), (\ref{K}), which thus determines the numerical factor $(1/2)^{1/3}$. In other models the factor $(1/2)^{1/3}$ would have to be replaced by some model dependent constant. The normalized cumulants of $F_{\text{GUE}}$ are \cite{B10.2}
\begin{align}
\label{C TW 1}
&\lim_{t\to\infty}C_{1}\approx-1.7710868074\,,\\
\label{C TW 2}
&\lim_{t\to\infty}C_{2}\approx0.8131947928\,,\\
\label{C TW 3}
&\lim_{t\to\infty}C_{2}^{-3/2}C_{3}\approx0.2240842036\,.
\end{align}
In Fig. \ref{fig cumulants} one observes how they are approached as $t \to \infty$. Most significantly, the mean decays as $t^{-1/3}$, implying that the Tracy-Widom distribution is approached from the left. The probability density function $F_{t}'(s)$ is plotted in Fig. \ref{fig Ft'} along with $F_{\text{GUE}}$. Note that the maximum first swings to the left and then to the right as dictated by the short time behavior and also reflected by $C_1$. 

The convergence to the Tracy-Widom distribution in the limit of long times has been established for a variety of discrete growth models. We refer to \cite{SS10.4} for a recent survey. $F_{\text{GUE}}(s)$ is the universal scaling function for droplet growth, i.e. for curved initial data. To next order, the difference between $F_{t}(s)$ and $F_{\text{GUE}}(s)$ scales as $t^{-1/3}$ \cite{SS10.2} when $t\to\infty$. The exponent $-1/3$ appears to be universal while the entire first order correction function is model-dependent \cite{FF11.1}.


\section{Conclusions}
We studied numerically the one-point height distribution function of the one-dimensional KPZ equation with sharp wedge initial data. Its universal part agrees very well with the experiment, in particular the long time convergence to the Tracy-Widom distribution and the relaxation of the mean as $t^{-1/3}$. The scheme developed by F. Bornemann turns out to be highly efficient and in essence reduces the computation to evaluating determinants from a 2-parameter family of $30 \times 30$ matrices. In a previous attempt \cite{SS10.2} regular grid approximation is used which requires much larger matrices, size $80 \times 80$, with accurate results only for times longer than $10$. Recently more exact solutions of the KPZ equation have become available, in particular the time-dependent joint height distribution at two spatial points at the same time for sharp wedge initial data \cite{PS11.1,PS11.2}, and the time-dependent height distribution at one point for flat initial data, $h(x,0) = 0$ \cite{CLD11.1}. Again the expressions involve Fredholm determinants, resp. Fredholm Pfaffians. It seems to us that the same numerical techniques used here could also be applied advantageously to these other cases.\bigskip\\
\textbf{Acknowledgements}. We are grateful to T. Sasamoto for a careful reading of the manuscript.


\end{document}